\documentclass[a4paper,12pt]{article}
\usepackage[font={footnotesize}]{caption}
\usepackage[T1]{fontenc}
\usepackage{xcolor}
\pdfoutput=1
 \usepackage{vmargin}            % redŽfinir les marges
\setmarginsrb{2cm}{2cm}{2cm}{2cm}{0cm}{0cm}{0cm}{0cm}
\usepackage{graphicx}\usepackage{amsmath}
\usepackage{amssymb,amsfonts,textcomp}
\usepackage{array}
\usepackage{supertabular}

\usepackage{hyperref}
\def\gtrsim{\lower.5ex\hbox{$\; \buildrel > \over \sim \;$}}
\usepackage{graphicx}

\newcommand{\hagn}{\mbox{{\sc \small Horizon-AGN\, }}}

\newcommand{\dd}{{\rm d}}
\newcommand{\be}{\begin{equation}}
\newcommand{\ee}{\end{equation}}

\begin{document}
\author{
Sandrine Codis\thanks{codis@cita.utoronto.ca}
\vspace*{6pt}\\
\small{Canadian Institute for Theoretical Astrophysics, University of Toronto},\\\small{
60 St. George Street, Toronto ON M5S 3H8, Canada}
}

\date{}
\title
{Nuisance parameters for large galaxy surveys}

\maketitle

\begin{abstract}
These notes are based on a lecture given at the 2016 Euclid Summer School in Narbonne.
I will first give a quick overview of the concept of nuisance parameters in the context of large galaxy surveys. The second part will examine the case study of intrinsic alignments, a potential important contamination of weak lensing observables.
\end{abstract}

%%%%%%%%%%%%%%%%%%%%%%%%%%%%%%%%%%%%%%%%%%%%%%%%%%%%%%%%%%%%%%%%%%%%%%%%%%
\section{Nuisance parameters in Bayesian analysis}
%%%%%%%%%%%%%%%%%%%%%%%%%%%%%%%%%%%%%%%%%%%%%%%%%%%%%%%%%%%%%%%%%%%%%%%%%%

\subsection{Statistical inference}

The Bayesian point of view interprets probabilities as a degree of belief\footnote{This is to be contrasted with the frequentist approach which has a different epistemological interpretation: probabilities are frequencies of occurence.}. 
Bayes' theorem  allows us to compute the probability of the parameters of a model given some data (statistical inference)
\begin{equation}
\label{eq:Bayes}
{\cal P}(\theta|d,m)=\frac{{\cal P}(d|\theta,m){\cal P}(\theta|m)}{{\cal P}(d|m)},
\end{equation}
where $m$ is the model, $\theta$ the set of parameters of the model and $d$ is the data. For instance, $m$ can represent the $\Lambda$CDM model, $\theta=\{\Omega_{m},\sigma_{8},\dots\}$ its parameters and $d$ is the observations: CMB temperature/polarisation maps, galaxy clustering, etc.
In Equation~(\ref{eq:Bayes}), ${\cal P}(\theta|d,m)$ is the so-called posterior distribution. It is the probability of the parameters of the model once the data is known, ${\cal P}(d|\theta,m)$ is the likelihood i.e the probability of the data given the parameters of the model, ${\cal P}(\theta|m)$ is the prior or in other words the probability of the parameters of the model before the data is known and ${\cal P}(d|m)$ is the Bayesian evidence (which is nothing but a normalisation).

Bayes' theorem is used in many fields, in particular in astrophysics and cosmology, to find not only the most likely values of the parameters of the model but their full statistics ${\cal P}(\theta|d,m)$ and therefore eventually draw figures of merit. In practice, one needs to
\begin{enumerate}
\item set the prior (do assumptions about the theory);
 \item compute the likelihood;
\item combine them and normalise to get the posterior distribution of the parameters.
\end{enumerate}
Step 1 is probably one of the most debated issues in Bayesian analysis.  If one wants to impose the least possible information on $\theta$ i.e use ignorance priors, it is straightforward to see that a flat prior will depend on the variable that is used ($\theta$ or $\log \theta$ or any other function of the parameters). However, iterative procedures if the initial prior is less informative than the likelihood, usually converges.

Step 2 is in general computationally very challenging and -- except for noticeable cases like low-dimension Gaussian settings -- it requires to rely on sampling of the posterior distribution (Markov Chain Monte Carlo, Approximate Bayesian Computation to name a few). Note that in many applications in cosmology, a Gaussian approximation by means of the Fisher matrix approach gives reasonable insights.

I propose a set of exercises in appendix~\ref{sec:exo} to play with Bayes' theorem.

\subsection{The concept of nuisance parameters}
The set of parameters $\theta=\{\alpha,\beta\}$ can be decomposed into two sets, the parameters of interest $\alpha$ (for instance the mass of neutrinos) on the one hand and the nuisance parameters $\beta$ (e.g galaxy bias) on the other hand.
In the Bayesian framework, one can consider $\alpha$ and $\beta$ as random variables and marginalize over $\beta$ to infer the statistics of the parameters of interest
\begin{equation}
{\cal P}(\alpha|d,m)=\int{\cal P}(\alpha,\beta|d,m) \dd\beta\,.
\end{equation}
Let us emphasize here that this is not equivalent at all to slicing the posterior at the most likely value for $\beta$. 

\begin{figure*}
\centering
\includegraphics[width=0.7\columnwidth]{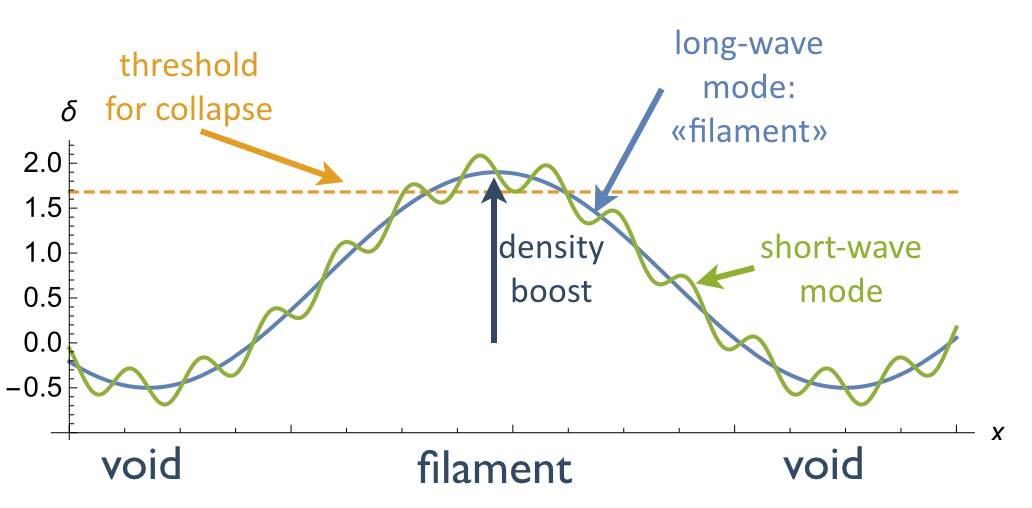} 
\caption{Peak Background split. A long wave fluctuation boosts smaller fluctuations and allows them to pass the threshold $\delta_{c}$ and collapse. The most massive objects are more likely to form in densest environments.
\label{fig:PBS}}
\end{figure*}

\subsection{Modeling the unknown with nuisance parameters}
\label{unkown}
In astrophysics, nuisance parameters can be used to model a wide range of unknown \footnote{Some examples were already highlighted in Francis Bernardeau and Martin Kilbinger's lectures} e.g from the instrument (calibrations, etc) or from a physical origin. Here I will briefly describe some of the physical unknowns that can be modeled using nuisance parameters.

\paragraph{Galaxy biasing}
Galaxies represent a biased tracer of the underlying matter density field. Therefore, when using galaxy clustering as a cosmological probe, one needs to model the so-called mass to light ratio or in other words galaxy biasing. It is commonly believed that most dark matter halos originate from the high peaks\footnote{Note that this is true only for the most massive objects. At lower mass, the shear becomes important and not all halos can be associated to a peak of the same mass \cite{2011MNRAS.413.1961L}} of the initial density field and the most massive the halo the highest the peak. This peak background split idea is illustrated in Figure~\ref{fig:PBS} and naturally induces a bias in the halo distribution that can be modeled by a local bias model \cite{2010PhRvD..81f3530G}.

\begin{figure*}
\centering
\includegraphics[width=0.7\columnwidth]{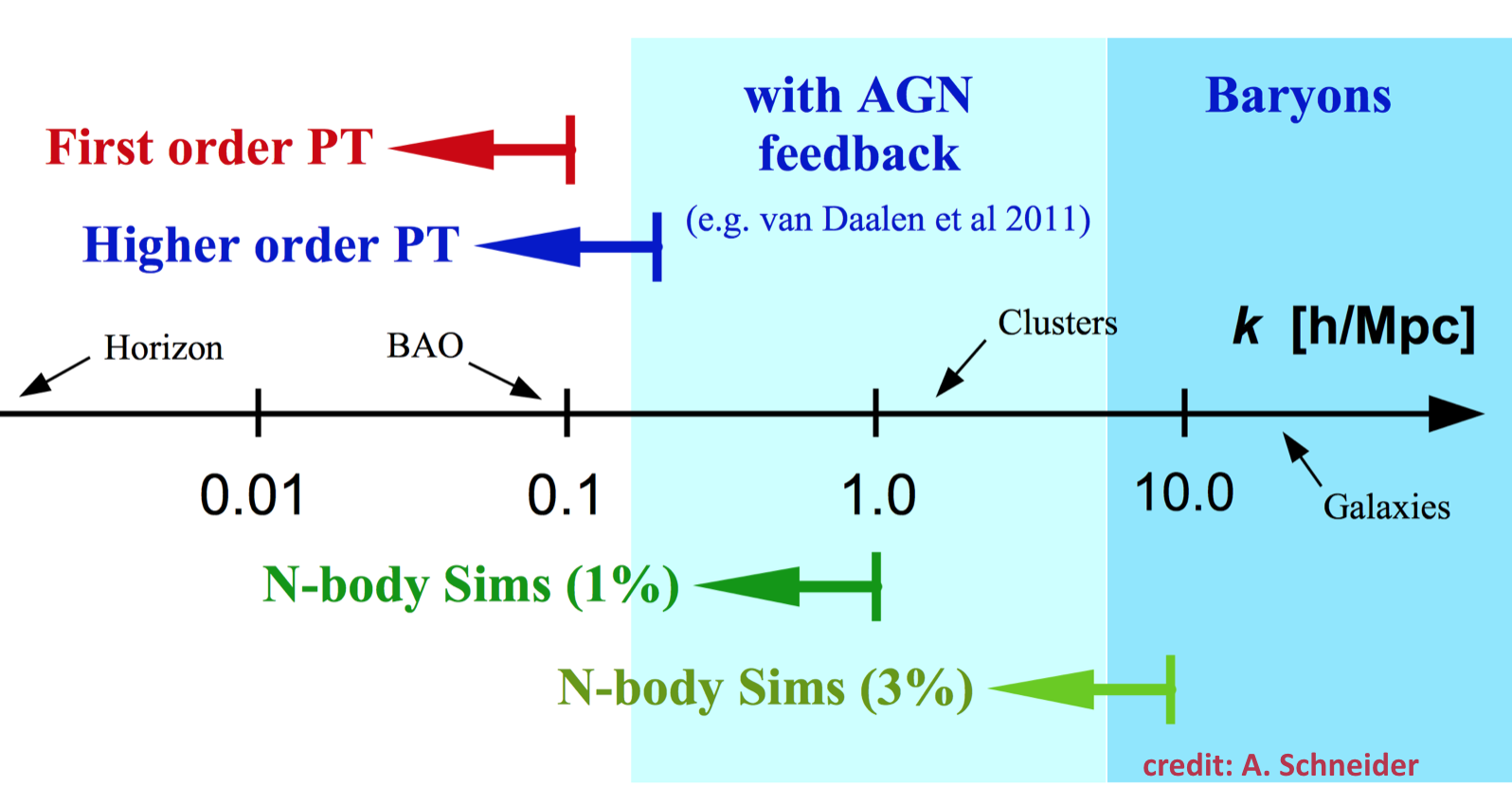} 
\caption{Accuracy of perturbation theory, N-body (pure dark matter) simulations and where baryons come into play (blue areas). Credit: Aurel Schneider.
\label{fig:baryons}}
\end{figure*}

 On relatively smaller scales, both deviation from the peak model and baryonic physics enter into play. In particular feedback from active galactic nuclei could have an impact on scales up to $0.15$Mpc$/h$ at low redshift \cite{2014MNRAS.440.2997V} (see also Figure~\ref{fig:baryons}). To model properly baryonic physics, hydrodynamical simulations are necessary but often not predictive because of the uncertainties on galaxy formation and therefore on subgrid recipes. The question is then : can we still parametrize baryonic physics and introduce nuisance parameters that encode our ignorance on the physics of galaxies? Shall we focus on the observables that are the least affected by baryons (e.g using PCA analysis \cite{2014MNRAS.440..240D})?

\begin{figure*}
\centering
\includegraphics[width=0.7\columnwidth]{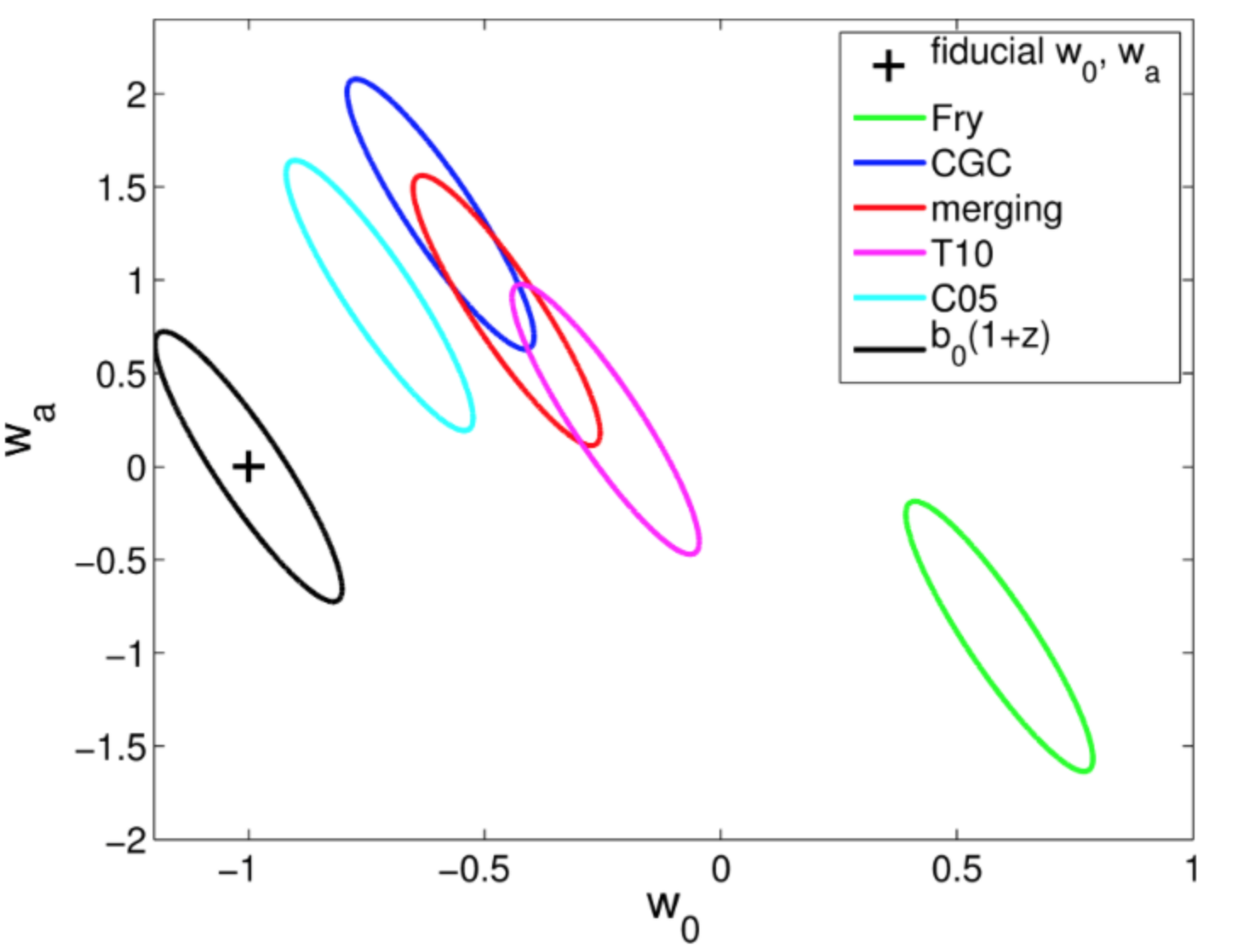} 
\caption{Impact of a wrong bias model $b(z)$ on the equation of state of dark energy $w(z)=w_{0}+w_{a}/(1+z)$. This figure is from \cite{2015MNRAS.448.1389C}.
\label{fig:bias}}
\end{figure*}

\paragraph{Redshift space distortions}
Galaxy surveys are conveyed in the so-called redshift space. Indeed, if the angular position of galaxies on the sky can be determined precisely, the radial distances usually rely on proxies for instance using the line-of-sight velocity component. Because galaxy velocities are not only due to the Hubble flow but also to their peculiar velocities, distortions appear in redshift space and are difficult to model because they couple strongly small and large scales. 
At the largest scales i.e in the linear regime of structure formation, the mapping from real to redshift space induces
an anisotropic change in the mass density contrast \cite{Kaiser87} which reads in Fourier space 
\begin{equation}
 \delta^{(s)}(\mathbf{k})=(1+f\mu^{2}) \delta^{(r)}(\mathbf{k}), 
\label{eq:Kaiser_mass}
\end{equation}
and depends on the angle $\mu={\bf k} \cdot {\bf \hat r}/k$
between the direction of ${\bf k}$ and the line of sight, and the amplitude, $f$, tracing
the growth history of linear inhomogeneities $D(a)$, 
$f= \dd \log D/\dd \log a \approx \Omega_{m}^{0.55}$ \cite{peebles80}. This Kaiser effect 
enhances the clustering by squeezing overdense regions and stretching underdense voids along the line of sight.
On small non-linear scales, fingers-of-God appear \cite{Jackson72,peebles80} stretching the collapsing clusters along the line-of-sight. 
Between those two regimes, a perturbative approach can be implemented  (e.g \cite{Scoccimarro99,bernardeau02} followed by the so-called streaming models of \cite{scoccimarro04} which include the finger-of-God effect). Unfortunately, because redshift space distortions introduce important couplings between small and large scales, a naive perturbation theory in redshift space has very bad convergence properties and can only describe scales exceeding several tens of megaparsecs. State-of-the-art approaches like the TNS model \cite{tns} can usually provide accurate predictions for the power spectrum for scales above $k\approx 0.2h/$Mpc when multiplying the extended Kaiser formula by a damping term
\begin{equation}
P(k,\mu)=D_{f}(k\mu f\sigma_{v}) \left[P_{\delta\delta}(k)+2f \mu^{2}P_{\delta\theta}(k)+f^{2}\mu^{4}P_{\theta\theta}(k)+A(k,\mu,f)+B(k,\mu,f)\right].
\end{equation}
The reader is refered to Francis Bernardeau's lecture for details about the modeling of redshift space distortions.

\paragraph{Non-linearities} If perturbation theory approaches can capture the first stages of structure formation, mode couplings quickly prevent us from having a parameter-free theory. Similar to the damping terms appearing for redshift-space distortions, the small-scale physics has eventually to be modeled by a set of nuisance parameters than can be either measured precisely in N-body simulations or marginalised over. As an illustration, here is the typical power spectrum at 2nd order obtained in the effective field theory approach \cite{2012arXiv1206.2926C} (similar results can also be found in other approaches like regPT\cite{2012arXiv1208.1191T}, time-sliced PT \cite{2016JCAP...07..028B}, etc).
\begin{equation}
P_{EFT}^{\rm 2-loop}=P_{11}+P_{\rm 1-loop}+P_{\rm 2-loop}-2(2\pi)(c_{s,1}^{2}+c_{s,2}^{2})\frac{k^{2}}{k_{\rm NL}^{2}}P_{11}+(2\pi)c_{s,1}^{2}P_{\rm 1-loop}+(2\pi)^{2}c_{s,1}^{4}\frac{k^{4}}{k_{\rm NL}^{4}}P_{11}.
\end{equation}
As the number of loops increases (the order of the perturbative expansion), the number of parameters like $c_{s,i}^{2}$ also increases. The question of the information gained by going more deeply into the non-linear regime but marginalising over an increased number of parameters is still an open issue.

\paragraph{Intrinsic alignments of galaxies} In weak lensing data, correlations of the intrinsic shapes of galaxies can contaminate the cosmological signal and need to be modeled. This topic will be developed further in the next section.

As a conclusion, for any cosmological observable, there are different sources of uncertainty and unknown which can be parametrised and marginalised over. This approach however has various drawbacks. For instance, the number of parameters (including nuisance) is huge and require high-dimensional and expensive statistical analysis.
Moreover, we can only get model-dependent constraints. In particular, in the era of precision cosmology, a slightly wrong model can significantly bias the resulting constraints on cosmological parameters. Because marginalising over the parameters of a given model does not guarantee that this model fits well the data, one also needs to do model comparison. 
Note that alternatives exist. For instance, one can use different parametrizations like the form filling function proposed by \cite{2009MNRAS.399.2107K}.
Having multiple observables with different biases and sensitivities is also key to get robust constraints on cosmology. For instance, besides the usual use of power spectrum and bispectrum, it will be important to also focus on alternative probes like extrema counts, Minkowski functionals, count-in-cell PDF, void profiles, etc.

\section{Intrinsic alignment: a nuisance for weak lensing probes}
\subsection{Cosmic shear contaminations}
Weak lensing is often presented as a potential powerful probe of cosmology for the coming years. It relies on the fact that the picture of galaxies that we observe in the sky is distorted because the light coming from background sources towards us is deflected by the gravitational potential wells along the line of sight. Therefore measuring these distortions can in principle constrain our cosmological model.
The idea is thus to try and detect coherent distortions of the shapes of galaxies e.g by using the two-point correlation function of the ellipticities of galaxies. But the apparent ellipticity of a galaxy $\epsilon$ comes from the cosmic shear $\gamma$  (which is related to the projected gravitational potential along the line of sight) but also from its intrinsic ellipticity $\epsilon_{s}$ (which actually accounts for almost $99\%$ of the apparent ellipticity). 
Therefore, the ellipticity-ellipticity two-point correlation function can be written as the sum of a shear-shear term but also contaminations
\begin{equation}
\left\langle\epsilon\epsilon'\right\rangle=\left\langle\gamma\gamma'\right\rangle+\left\langle\gamma\epsilon'\right\rangle+\left\langle\epsilon\gamma'\right\rangle+\left\langle\epsilon\epsilon'\right\rangle
\end{equation}
where the prime denotes another galaxy at a given angular distance. These contributions that somehow contaminate the shear signal are the so-called intrinsic alignements, $\left\langle\gamma\epsilon'\right\rangle+\left\langle\epsilon\gamma'\right\rangle$ being the GI contribution and $\left\langle\epsilon\epsilon'\right\rangle$ the II contamination. 
A priori, the ellipticity-ellipticity two-point correlation function is expected to be dominated by the cosmic shear signal. Indeed,
one could expect the intrinsic alignments to be negligible because the shape of galaxies seems to be uniformly distributed in the Universe. However, with future surveys claiming percent precision on the equation of state of dark energy, this rough approximation has to be revisited because galaxies are believed to be correlated with the cosmic web. The consequence of this large-scale coherence of galaxies could then contaminate significantly weak lensing observables, at a level of about ten pourcents.
In section~\ref{sec:phys}, I will describe how galaxies seem to be indeed correlated to their large-scale filamentary environment.
I will then describe the theoretical models of intrinsic alignments in section~\ref{sec:NLA} before showing results obtained recently on hydrodynamical simulations in section~\ref{sec:hydro}.

\subsection{Physical origin}
\label{sec:phys}
\begin{figure*}
\centering
\includegraphics[width=0.5\columnwidth]{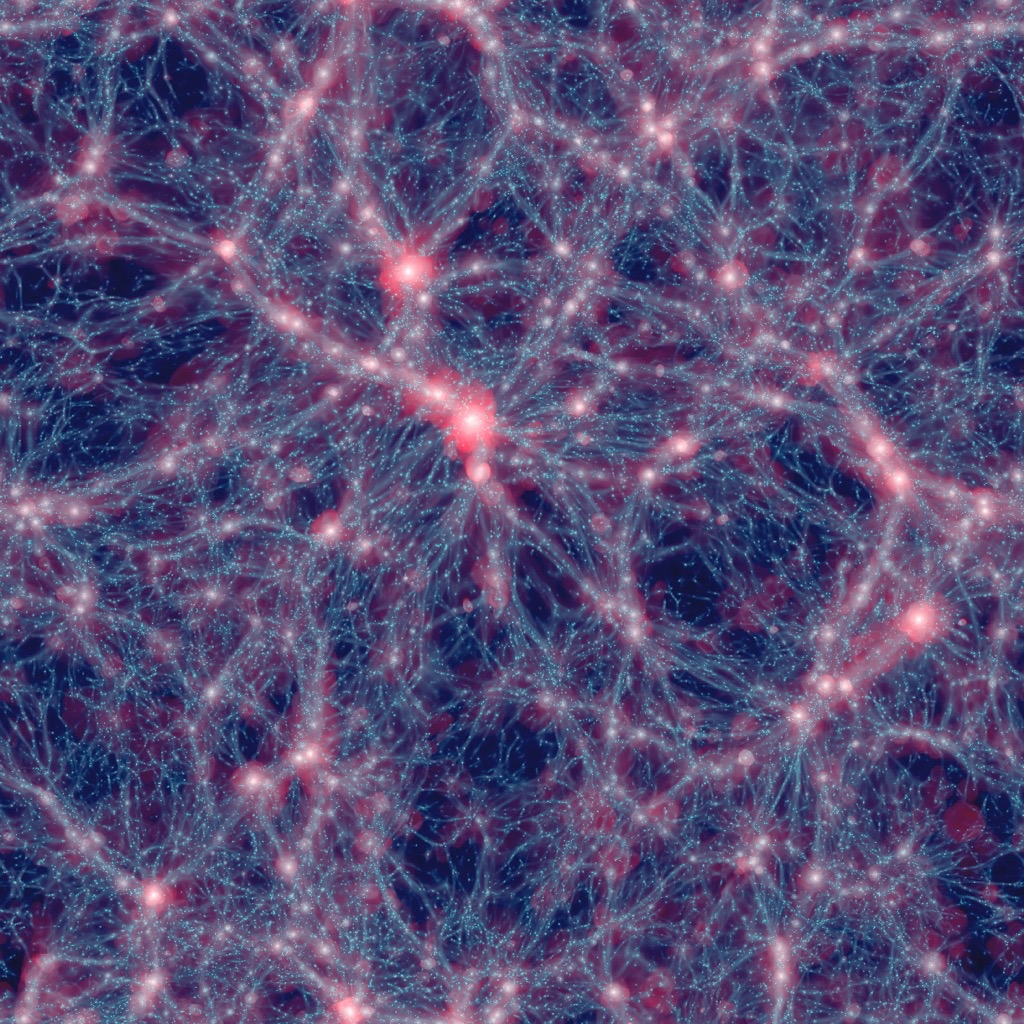} 
\caption{Filaments of the Horizon-AGN simulation \cite{dubois14}.
\label{fig:LSS}}
\end{figure*}
The large-scale structure of the Universe is the birthplace of galaxies. This filamentary structure is seen on Figure~\ref{fig:LSS}. If galaxies form and evolve within these large cosmic highways, a natural question which arises is to understand the role played by this anisotropic environment on galaxy formation, evolution and properties.
It is now well-established that the morphology of galaxies is not independent from the local density: elliptical galaxies are found preferentially in dense regions and spiral galaxies are found in the field \cite{oemler74,dressler80,hermit96,guzzo97}, probably as a result of the peak background split discussed in section~\ref{unkown}. Other galaxy properties are also found to be correlated to their environment, for instance colour, luminosity, surface brightness or spectral type \cite{norberg02,hogg03,blanton05,LeeLee08,LeeLi08,tempel11,yan12,kovac14}. Beyond a purely scalar effect, 
observations also indicate that the rotation axes of galaxies are correlated with the filaments in which they are embedded \cite{navarro04,trujilloetal06,leeetal07,pazetal08,jones10,Tempel13,zhangetal13,tempel&libeskind13}.

\begin{figure*}
\centering
\includegraphics[width=1\columnwidth]{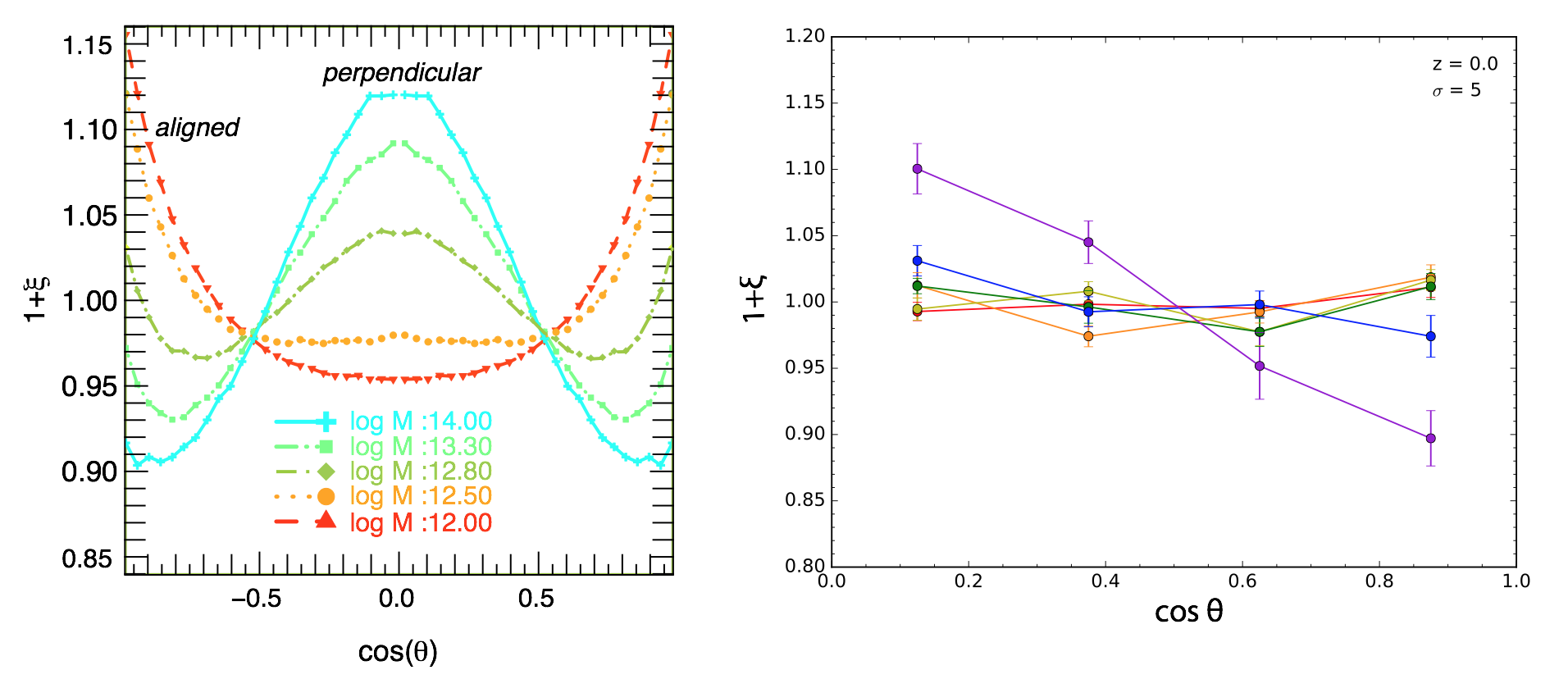} 
\caption{PDF of the cosine of the angle between the spin and the closest filament at redshift 0. Left-hand panel: results obtained from the 73 millions dark matter halos of the Horizon-4$\pi$ run for masses between $10^{12}$ and $10^{14} M_{\odot}$ as labeled \cite{codisetal12}. Right-hand panel: results obtained from the virtual galaxies of the Horizon-AGN simulation for masses between $10^{8.5}$ and $10^{11} M_{\odot}$ from red to violet (Jindal et al, in prep.).
\label{fig:spin}}
\end{figure*}

Studying the properties of galaxies as a function of their location in the cosmic web yields valuable information about the formation and evolution of galaxies. It is therefore crucial to use numerical simulations to investigate how virtual galaxies and halos are correlated to their large-scale environment.
\cite{hahnetal07,hahnetal07b,hahn09,gayetal10,metuki15} found that the properties of dark matter halos such as their morphology, luminosity, colour and spin parameter depend on their environment as traced by the local density, velocity and tidal field. Spin-1 and 2 observables also display correlations with the cosmic web.  For instance, their shape and spin are correlated to the directions of the surrounding filaments and walls both in dark matter \cite{aubertetal04,bailin&steinmetz05,brunino07,hahnetal07,hahnetal07b,calvoetal07,pazetal08,zhangetal09,codisetal12,libeskind13,aragon14} and hydrodynamical simulations \cite{navarro04,hahn10,dubois14}. Massive halos have a spin preferentially perpendicular to the axis of their host filament while small-mass halos tend to have a spin aligned with the closest filament as displayed on Figure~\ref{fig:spin}. The transition between those two regimes is redshift-dependent and slightly scale-dependent. A similar behaviour is also detected for virtual galaxies (\cite{dubois14}, Jindal et al, in prep.) namely massive, red, elliptical galaxies are more likely to have a spin perpendicular to the axis of the filament whereas the intrinsic angular momentum of small-mass, blue, star forming and disk-like galaxies tends to be parallel to the filamentary structure nearby. These predictions agree with recent observations for instance in the SDSS by \cite{Tempel13}.

But how to understand this phase transition in the spin-filament alignment from aligned at low mass to perpendicular at large mass? In the early stages of structure formation, matter flows inside the walls towards the forming filament. This process generates vorticity aligned with the axis of the filament as shown by \cite{pichon99,laigle2014,libeskind13}. The first generation of small halos forms within these vorticity-rich filaments and naturally acquires spin also aligned with the axis of the filament. Later, matter streams inside the filaments towards the nodes of the cosmic web, some halos catch up with each other, merge, transform their orbital angular momentum into spin and get more massive. A second generation of halos, more massive, is born with spin perpendicular to the direction of motion i.e perpendicular to the filaments. The more numerous the mergers, the more perpendicular the spin \cite{welker2014}.

Interestingly, this spin flip can also be understood from first principles in a Lagrangian framework \cite{ATTT} where spins are acquired by tidal torquing and tides depend on the location in the cosmic web. The basic ideas behind tidal torque theory will be exposed in section~\ref{sec:NLA}.

\subsection{Theoretical modeling of intrinsic alignments}
\label{sec:NLA}
If galaxies are correlated with the cosmic web as I showed in section~\ref{sec:phys}, it is natural to expect intrinsic alignments. Two different mechanisms are commonly referred to in this context, one for disc-like galaxies, one for ellipticals.

Discs are dominated by their spin which is believed to be generated at their formation time by tidal torquing. This leads to the so-called quadratic alignment model. At linear order, tidal torque theory states that the spin is acquired gradually until the time of maximal extension (before collapse) and is proportional to the misalignment between the inertia tensor of the protogalaxy and the tidal tensor due to the surrounding distribution of matter (see \cite{schaefer09} for a review)
\begin{equation}
\label{eq:TTT}
L_{i}=a^{2}(t)\dot D_{+}(t)\sum_{j,k,l}\epsilon_{ijk} I_{jl}T_{lk}\,,
\end{equation}
where $a(t)$ is the scale factor, $D_{+}$ the growth factor, $T_{ij}$ the tidal tensor (detraced Hessian of the gravitational potential) and $I_{ij}$ the protogalactic inertia tensor. If the shape of the proto-halo is spherical or if its inertia tensor is aligned with the tidal tensor then no spin is acquired at first order. A quick derivation of this result is proposed in Appendix~\ref{app:TTT}.
Once the spin is modeled by Equation~(\ref{eq:TTT}), one can use a thin disk approximation and deduce the expected shape correlation functions as was done e.g by \cite{Lee01,Cri++01}.

\begin{figure*}
\centering
\includegraphics[width=0.8\columnwidth]{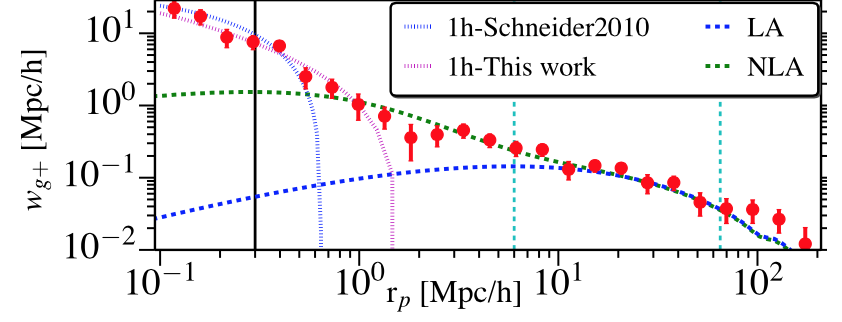} 
\caption{Projected two-point correlation between galaxy positions and galaxy shapes in the SDSS LOWZ sample \cite{2015MNRAS.450.2195S}. Measurements in red are compared to the linear alignment model (LA), non-linear alignment model (NLA) and halo models (dotted line). 
\label{fig:LRG}}
\end{figure*}

On the other hand, the star distribution in elliptical galaxies is stretched by 
the external tidal field which leads to a linear alignment model \cite{Cat++01,H+S04}. Here, the intrinsic ellipticity  is proportional to the tidal field, i.e second derivatives of the gravitational potential smoothed on some scale $\Psi$, with a coefficient $C_{1}$ that measures the strength of the alignment occurring at redshift $z_{\textrm{IA}}$
\begin{equation}
\label{eq:linearmodel}
 \left(\begin{array}{c}
e_{s}^{+}\\
e_{s}^{\times}\\
\end{array}
\right)
=-\frac{C_{1}}{4\pi G}
 \left(\begin{array}{c}
\nabla_{x}^{2}-\nabla_{y}^{2}\\
2\nabla_{x}\nabla_{y}\\
\end{array}
\right)
\Psi (z_{\textrm{IA}})\,,
\end{equation}
as it is assumed that stars and dark matter are in dynamical equilibrium and therefore the stellar distribution follows the distortion of the galaxy halo spheroid which is tidally distorted.
The strength coefficient was for instance measured in the SuperCOSMOS field $C_{1}\approx 5\cdot 10^{-14}(h^{2}{\textrm{ M}}_{\odot}{\textrm{Mpc}})^{-1}$ \cite{B+K07}.
This ansatz can be used to predict the GI and II signals for a linearly evolved potential field.
For instance, the two-point correlation function between density and ellipticity reads in Fourier space
\begin{equation}
  P_{g+}({\bf k},z) = -\frac{b_gC_1\rho_{\rm crit}\Omega_m}{D(z)}\frac{k_x^2-k_y^2}{k^2}P_\delta({\bf k},z)\,.
  \label{eq:powerdi}
\end{equation}
If this model seems accurate on scales above 10 Mpc/h \cite{blazek11}, it can be significantly improved on smaller scales by taking into account the non-linear dark matter evolution \cite{B+K07} and other non-linear contributions like non-linear bias and the effect of weighting by the local density of galaxies \cite{blazek15}. A halo model can also be implemented to improve the small-scale modeling \cite{S+B10}. 
Figure~\ref{fig:LRG} shows the alignment of luminous red galaxies in the SDSS \cite{2015MNRAS.450.2195S} and compares with linear, non-linear alignment models and halo model. A combination of halo model on small scales and non-linear alignment model on large scales (NLA+halo) was found to accurately fit the data.
At present day, no detection of blue galaxies intrinsic alignments has been reported \cite{Man++11}.

\subsection{Intrinsic alignments in hydrodynamical simulations}
\label{sec:hydro}
While understanding intrinsic alignments from first principles can give interesting highlights on large scales, it seems however necessary to combine these analytical models with numerical simulations in order to model intrinsic alignments in the strong non-linear regime where the effect of baryonic physics also becomes important.

Early investigations focused on dark matter only simulations assuming that galaxy shapes were following dark halo shapes \cite{HRH00}. Then halo model and semi-analytical modeling have been used to assign galaxies to their host halo \cite{Joa++13b}. It is only recently that hydrodynamical simulations have been able to describe a sufficient large volume with enough resolution so as to measure how numerical galaxies were intrinsically aligned \cite{codis14,Tenneti15a,chisari15,Velliscig15b,2016MNRAS.461.2702C}. Table~\ref{table} sums up those recent simulations that have been used to study intrinsic alignments. They use different codes (based on particles or grid), different boxsize, resolution and subgrid physics including cooling, star formation and both supernovae and AGN feedback to resolve respectively the low-mass and high-mass end discrepancy in the galaxy mass function.

\begin{table}
\centering
\begin{tabular}{|c|c|c|c|c|c|}\hline
simulation & code & paper & box size (Mpc$/h$) & resolution\\\hline
Horizon-AGN & RAMSES (adaptative mesh) & \cite{dubois14} & 100 & $1024^{3}$ \\\hline
Cosmo-OWLS/Eagle & GADGET-3 (particles)  &\cite{eagle} & 100 & $1504^{3}$ \\\hline
Illustris & AREPO (moving mesh) &  \cite{illustris} & 75 & $1820^{3}$ \\\hline
MassiveBlack II & P-GADGET (particles) & \cite{massiveblack} & 100&$1792^{3}$\\\hline
\end{tabular}
\caption{Cosmological hydrodynamical simulations that have been used to measure intrinsic alignments.}
\label{table}
\end{table} 

\begin{figure*}
\centering
\includegraphics[width=0.6\columnwidth]{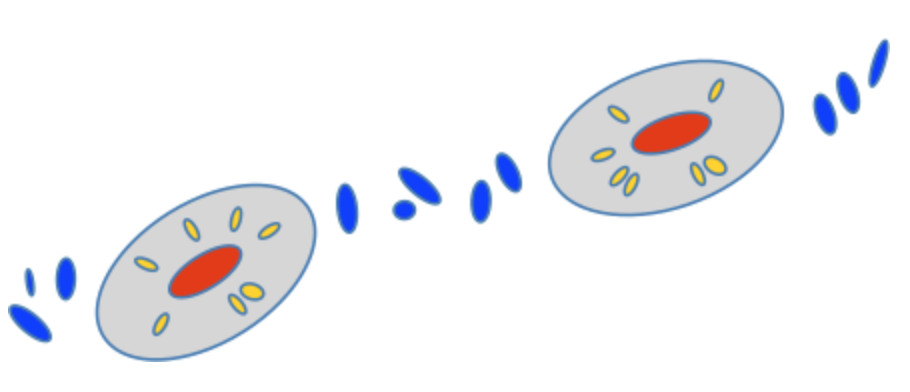} 
\caption{Sketch of the alignments of spirals (in blue) and ellipticals (in red) with the cosmic web \cite{chisari15}.
\label{fig:cartoon}}
\end{figure*}

\begin{figure*}
\centering
\includegraphics[width=0.6\columnwidth]{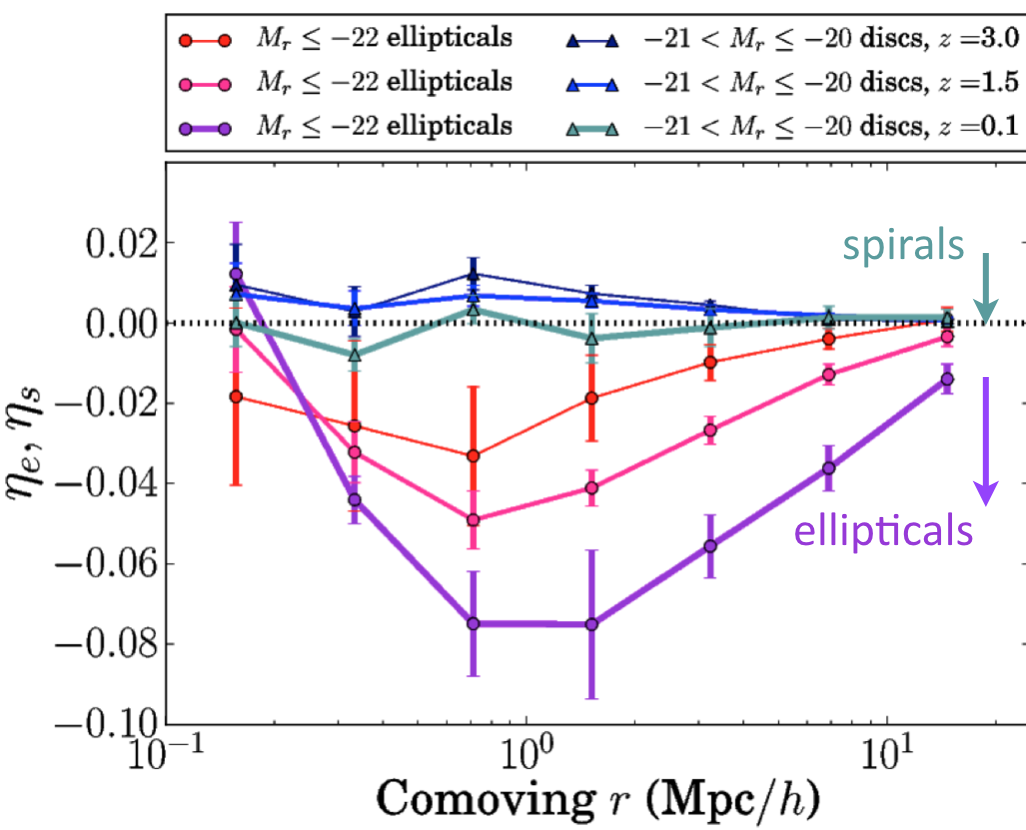} 
\caption{Alignment of galaxy shapes with the large-scale structure for different morphologies and redshifts as labeled \cite{2016MNRAS.461.2702C}. A negative value means radial alignment while a positive value stands for a tangential alignment.
\label{fig:align-z}}
\end{figure*}

The Horizon-AGN simulation has been extensively used to study intrinsic alignments of galaxies. First investigation by \cite{codis14} focused on the II contamination by measuring the spin-spin correlation function of the 165 000 virtual galaxies at z=1.2 and converting the signal in a projected ellipticity-ellipticity correlation using a thin disk model. They found a clear signal for blue galaxies on scales below 10 arcseconds but no signal for red galaxies because the spin is a bad proxy for their shape. Later, \cite{chisari15} used the inertia tensor to define the ellipticity of galaxies and studied the GI alignments using the so-called orientation-separation correlation function which measures the correlation between the shape of galaxies and the density field. The advantage of this estimator is twofold: first this is often the observable used with real data, second, it does not suffer from grid locking. \cite{chisari15} found that ellipticals are elongated radially towards all galaxies and spirals are elongated tangentially w.r.t ellipticals as seen on the cartoon displayed in Figure~\ref{fig:cartoon}. In projection, only the signal from ellipticals pervades and was shown to agree well with observations and NLA+halo model. This signal evolves with redshift: \cite{2016MNRAS.461.2702C} showed that the radial alignment of ellipticals increase at low redshift while the tangential alignment of disks increases at high redshift and could become an important source of contamination for future surveys as can be seen on Figure~\ref{fig:align-z}. The amplitude of the contamination also depends on the way shapes are measured.

\begin{figure*}
\centering
\includegraphics[width=0.8\columnwidth]{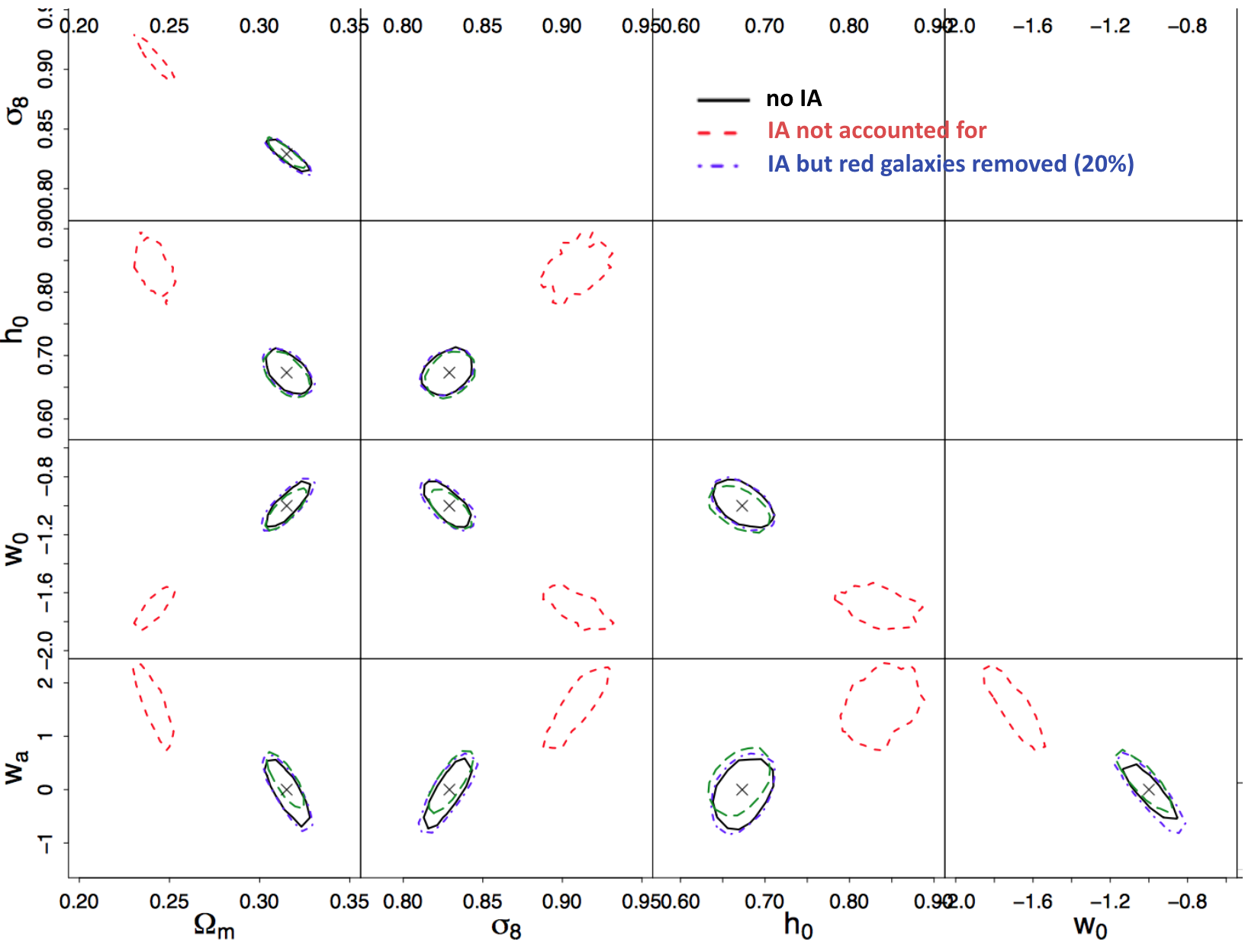} 
\caption{Impact of intrinsic alignments on Euclid cosmological constraints  \cite{2016MNRAS.456..207K}. Black contours correspond to the case when there is no intrinsic alignment, red when there are intrinsic alignments but they are not taken into account and blue when red galaxies are removed (as they are the most affected by intrinsic alignments).
\label{fig:Krause}}
\end{figure*}

However, a number of issues remain. All simulations do not completely agree. For instance, Illustris and Massive-Black do not predict the tangential alignment of spiral galaxies and the increased radial alignment at low-redshift \cite{2016MNRAS.462.2668T}. Theoretical models can not account for all measurements performed in \hagn, for instance they fail at reproducing the redshift and luminosity dependence.
Understanding the impact of baryonic physics, numerical schemes, adding observational systematics like dust attenuation, shape measurements, galaxy selections would be crucial for future surveys as intrinsic alignments can have a drastic impact on cosmological constraints obtained from weak lensing data (see Figure~\ref{fig:Krause}). Two main strategies can be used : use a parametric model accurate enough and marginalise over these nuisance parameters or select a population of galaxies less prone to intrinsic alignments but as we have discussed, blue galaxies may not be safe at higher redshift as their alignments increase. Because the intrinsic alignments correlate with the cosmic web, combining ingeniously weak lensing data with galaxy clustering may be a good way to mitigate the nuisance of intrinsic alignments.

\section{Conclusion}
\label{sec:conclusion}

Galaxies are correlated to the large-scale structure which induces intrinsic alignments and can contaminate weak lensing observables. Parametrizing this effect is challenging as it requires to model properly the small-scale physics including the highly non-linear regime and baryonic physics. Additional works both from the numerical and the theoretical sides are needed to reach the goal of percent precision on cosmological parameters set for future surveys. Finding the optimal strategy to mitigate the effect of intrinsic alignments is still an open issue.

{\scriptsize
\bibliographystyle{plain}
\bibliography{references}

}
\appendix
\section{Tidal torque theory in a nutshell}
\label{app:TTT}

\begin{figure}
\begin{center}
\includegraphics[width=0.9\columnwidth]{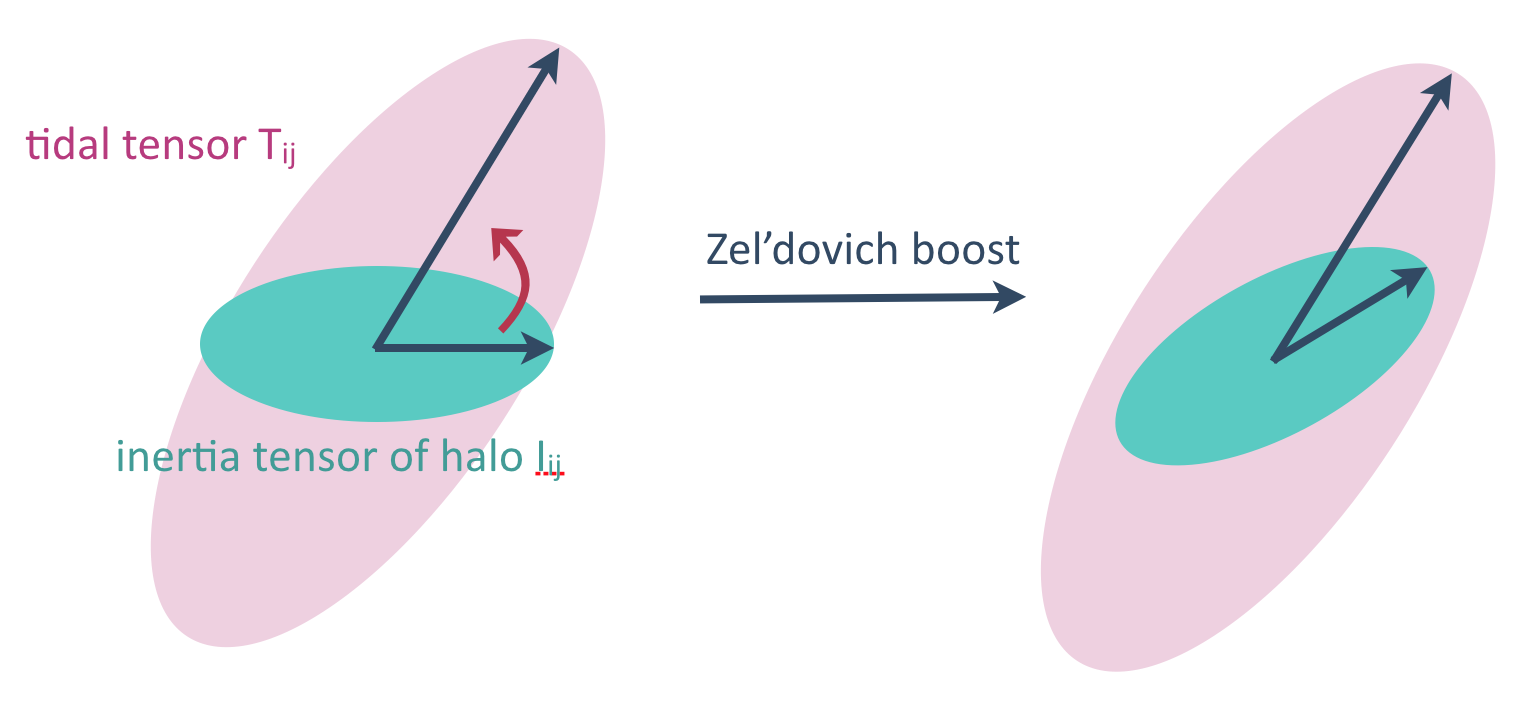}
\caption{Spin
acquisition by tidal torquing. At linear order, the misalignment between the inertia tensor of the proto-object and the surrounding tidal tensor induces an inhomogeneous Zel'dovich boost which corresponds to the acquisition of a net intrinsic angular momentum in Eulerian space.
\label{fig:TTT}}
\end{center}
\end{figure}

In the standard paradigm of galaxy formation, protogalaxies acquire their spin by tidal torquing coming from the surrounding matter distribution \cite{hoyle49,peebles69,doroshkevich70,white84,catelan96,Cri++01}. At linear order, this spin is acquired via Equation~(\ref{eq:TTT}).
Indeed, the spin of a protogalaxy contained in a volume V and with center of gravity located at position 
$\bar{\mathbf{r}}$  can be written
\begin{equation}
\label{eq:Ldef}
\mathbf{L}=\int_{V}\dd^{3} \mathbf{r} (\mathbf{r}-\bar{\mathbf{r}}) \times (v(\mathbf{r})-v(\bar{\mathbf{r}}))\rho(\mathbf{r})\,,
\end{equation}
where the implicitly time-dependent velocity field is denoted $v(\mathbf{r})$ and the mass density $\rho(\mathbf{r})$. In Lagrangian coordinates, Equation~(\ref{eq:Ldef}) becomes
\begin{equation}
\label{eq:Llag}
\mathbf{L}=\rho_{0}a^{5}\int_{V_{L}}
\dd^{3} \mathbf{q} (\mathbf{x}-\bar{\mathbf{x}}) 
\times (\dot{ \mathbf{x}}(\mathbf{q}) -\dot{ \mathbf{x}}(\bar{\mathbf{q}}) )\,.
\end{equation}
In the Zel'dovich approximation where $\mathbf{x}=\mathbf{q}-D_{+}\nabla\psi(\mathbf{q})$ and $\dot {\mathbf{x}}=-\dot D_{+} \nabla \psi$, $\psi $ being the displacement field (such that $\Delta \psi=\delta$), and assuming that the gradient of the displacement field is almost constant across the proto-object of Lagrangian volume $V_{L}$, Equation~(\ref{eq:Llag}) can be evaluated by a second-order Taylor expansion
\begin{equation}
\mathbf{L}\approx-\dot D_{+}\rho_{0}a^{5}\int_{V_{L}}\dd^{3} \mathbf{q} (\mathbf{q}-\bar{\mathbf{q}}) \times \left[T\cdot(\mathbf{q}-\bar{\mathbf{q}})\right] \,,
\end{equation}
where $T$ is the tidal shear tensor $T_{ij}=\partial_{i}\partial_{j}\Psi_{ij}$ at the center of gravity.
Let us define the inertia tensor $I$
\begin{equation}
I_{ij}=\rho_{0}a^{3}\int_{V_{L}}\dd^{3} \mathbf{q} ({q}_{i}-\bar{{q}}_{i})({q}_{j}-\bar{q}_{j})\,,
\end{equation}
so that the spin of the proto-galaxy can eventually be written as Equation~(\ref{eq:TTT}). It is mainly advected until the time of maximal extension before collapse as the lever arm is then drastically reduced. 
This process of spin acquisition by tidal torquing is illustrated on figure~\ref{fig:TTT}. It is clear from equation~\ref{eq:TTT} that spherical proto-objects can not acquire angular momentum in this context but it can be shown \cite{peebles69} that spin acquisition in spherically symmetric settings is possible as a second-order effect.

In the Lagrangian picture, $I_{ij}$ is the moment of inertia of a uniform
mass distribution within the Lagrangian image of the halo, while $T_{ij}$ is
the tidal tensor averaged within the same image. Thus, to rigorously determine
the spin of a halo, one must know the area from which  matter is assembled,
beyond the spherical approximation. While this can be determined in numerical
experiments, theoretically we do not have the knowledge of the exact boundary
of a protohalo. As such, one inevitably 
has to introduce an approximate proxy for 
the moment of inertia (and an approximation for how  the tidal field is averaged over that region).

The most natural approach is to consider that protohaloes form around an elliptical
peak in the initial density and approximate its Lagrangian boundary with
the elliptical surface where the over-density drops to zero. 
This leads to the following 
approximation for the inertia tensor \cite{catelan96,Schafer2012}
\begin{equation}
\label{Iij}
I_{ij}=\frac M 5  \left(
\begin{array}{ccc}
A_{y}^{2}+A_{z}^{2}&0&0\\
0&A_{z}^{2}+A_{x}^{2}&0\\
0&0&A_{x}^{2}+A_{y}^{2}
\end{array}
\right)\,,
\end{equation}
(in the frame of the Hessian) where the mass of the protohalo is $M=4/3 \pi A_{x}A_{y}A_{z}\rho_{0}a_{0}^{3}$ and the semi-axes of the ellipsoid, $A_i$,  are function of the eigenvalues of the Hessian (negative for a peak),
\begin{equation}
A_{i}=\sqrt{\frac{2\nu\sigma_{2}}{-\lambda_{i}}}\,.
\end{equation}
with $\nu$ the peak overdensity.
Note that the traceless part of $I_{ij}$ that is relevant for torques is then
proportional to the traceless part of the inverse Hessian of the density field
\begin{equation}
\label{eq:inertia}
\overline{I}_{ij} = \frac{2}{5} \nu \sigma_2 M \overline{H}_{ij}^{-1}\,.
\end{equation}

This scenario of spin acquisition by linear tidal torquing has been tested against numerical simulations. It appears that the spin (in particular its direction) is well-predicted in the early stages of galaxy evolution ($z<3$) before non-linearities beyond Zel'dovich become important \cite{ej79,be87,sugermanetal00,porcianietal02,porcianietal02b,Lee08}. In particular, they confirm two predictions of linear tidal torque theory : the spin scales like the mass to the power $5/3$ and the spin parameter $\lambda=L/\sqrt{GRM^{3}}$ is anti-correlated with the peak height (the higher the peak, the lower the spin parameter). However at later times, both spin direction and magnitude deviate significantly from the linear prediction.

\section{Exercises : Bayesian analysis and nuisance parameters}
\label{sec:exo}
\subsection{The Gaussian case}
Let's assume that we have a cosmological probe with a Gaussian likelihood of mean $\theta_{m}$
 and covariance matrix $C$ and that the prior is also Gaussian centred on the maximum likelihood with covariance matrix $\Sigma$.

b) Demonstrate that the posterior distribution is also Gaussian and compute the mean and covariance matrix.

c) Generalize the result in the case of multiple probes.

\subsection{Modelling biased tracers}
We want to estimate the dark matter variance ($\sigma$) from measurements of the density of biased tracers (halos, galaxies) in spheres of a given radius $R$. 
Let's
 assume we have a model for the PDF of dark matter density contrasts in spheres of radius R which depends on one parameter: $\sigma$. This PDF is called $P[\sigma](\delta)$.
 
 d) Compute the PDF of the density of biased tracers $P_{h}[\sigma,b](\delta_{h})$ if we assume a linear bias relation :  $\delta_{h}= b \delta$. 
 
We now have a two-parameter model. In what follows, we will assume that $P$ is a lognormal distribution (meaning that $\ln (1+\delta)$ follows a Gaussian distribution)\footnote{Note that it is possible to predict this density PDF beyond the lognormal approximation with the so called large deviation approach (see for instance \url{http://cita.utoronto.ca/~codis/LSSFast.html})}.

e) Compute the variance and mean of the PDF of $\ln (1+\delta)$ in order for $\delta$ to have zero mean and variance $\sigma$.

1000 random numbers that follow the distribution $P_{h}[\sigma_{0},b_{0}]$ have been generated and are stored in the file ``NuisanceParameters-data.dat'' available here \url{https://dl.dropboxusercontent.com/u/4958235/NuisanceParameters-data.dat}.

f) Assuming flat priors, compute the two-dimensional posterior distribution. Plot it using three contours corresponding to the 1,2,3 sigma regions. What is the most likely value of $(\hat\sigma,\hat b)$?  b is a nuisance parameter, what is the constraint you can put on $\sigma$ ?

g) Redo the same analysis for a quadratic bias model $\delta_{h}= b_{1} \delta+b_{2} \delta^{2}$ (\url{https://dl.dropboxusercontent.com/u/4958235/NuisanceParameters-data-quadratic.dat}). What happens if you assume a linear model to fit the data? Compute the Bayes factor (ratio of evidences) of the linear and quadratic models in that case.

\end{document}